# EFFECTS AND PROPOSITIONS

*William Demopoulos*[*]

## Abstract

The quantum logical and quantum information-theoretic traditions have exerted an especially powerful influence on Bub's thinking about the conceptual foundations of quantum mechanics. This paper discusses both the quantum logical and information-theoretic traditions from the point of view of their representational frameworks. I argue that it is at this level—at the level of its framework—that the quantum logical tradition has retained its centrality to Bub's thought. It is further argued that there is implicit in the quantum information-theoretic tradition a set of ideas that mark a genuinely new alternative to the framework of quantum logic. These ideas are of considerable interest for the philosophy of quantum mechanics, a claim which I defend with an extended discussion of their application to our understanding of the philosophical significance of the no hidden variable theorem of Kochen and Specker.

## 0. Introduction

I should begin with two remarks about the nature and scope of this study. The first remark concerns my use of the notion of an effect and its juxtaposition with the concept of a proposition in my title. The connection between propositions and effects is a subject about which I will have a great deal to say in the course of the development of my positive proposals. But to forestall possible misunderstandings, and as a precaution against raising false expectations, I want to say at the outset that this paper does not address the similarities and differences between effect algebras and orthomodular posets of propositions; nor does it concern positive operator valued measures (POVMs) and their relation to projection valued measures (PVMs). Indeed, I make no use of the notion of an effect algebra, or that of a POVM. My concerns are philosophical and conceptual rather than mathematical and foundational. The paper neither contains nor promises results of the sort one might expect of a study of effect algebras and their relation to the axiomatic tradition of quantum logic. It should go without saying that by ignoring effect algebras and POVMs, I do not imply a judgement about their utility, interest or importance for the foundations or philosophy of the subject; it is merely that my interests lie elsewhere. Unfortunately the term 'effect,' although already well-established with a technical meaning, happens to suit my purposes almost perfectly, and I have been unable to find an alternative that works as well.

---

[*] I have been influenced in more ways than I have been able to record in the text by my friends and colleagues, Jeffrey Bub, Itamar Pitowsky, Robert Di Salle, and most recently, Christopher A. Fuchs, My thanks to them for many hours of conversation and many pages of correspondence. My research was supported by the Social Sciences and Humanities Research Council of Canada.

The source of my interest in the concept of an effect, in the sense in which it occurs here, is the result of reflection on conceptual aspects of the work of Fuchs and his collaborators; as I came to realize only later, it also depends on an idea implicit in the work of Pitowsky (especially his 2003 and 2006). As will become clear in the sequel, the central mathematical concept on which this paper rests is familiar from the tradition of axiomatic quantum logic. Effects and propositions are investigated for their suitability as the objects which, from the point of view of the conceptual foundations of the theory, most naturally realize this mathematical structure.

My second remark concerns the scope of the discussion to follow. There are at least three groups of conceptual issues associated with the interpretation of quantum mechanics. There is first the cluster of "puzzling phenomena" permitted by the theory. The most famous among these are the 2-slit experiment, the EPR thought experiment, and the latter's striking realization in work carried out in response to Bell's (1964) discoveries. Secondly, there is the problem, variously formulated, of reconciling the apparent conflict between the discontinuous state transition, which occurs in the context of a measurement, and the continuous evolution to which a quantum mechanical system conforms in the absence of measurement. Thirdly, there is the quantum-theoretical possibility of a finite family of properties associated with a physical system that are so interrelated that there is no 0-1 generalized probability measure definable on them.

Following tradition, I call the puzzling phenomena which comprise the first group, quantum paradoxes, and the second, the measurement problem. I depart from tradition by calling the problem of hidden variables the essential content of one of the two principal theorems which are often invoked as a *solution* to what usually passes for this problem;

more precisely, by *the problem of hidden variables* I mean the problem of understanding the significance of the possibility of such finite families of properties.

I believe this is a useful division of problems even though there are evidently connections among the issues each of them raises. Thus, to take just one example, after Bell's analysis, no discussion of hidden variables would be complete if it did not address EPR and issues connected with locality, non-separability and entanglement; and the original point of each of the paradoxes I have mentioned was to highlight some aspect of the measurement problem or a difficulty with one or another proposed solution of it: EPR addressed naïve disturbance theories of measurement, and the 2-slit paradox is, among other things, an early illustration of the difficulties measurement presents. I claim that the classification is nevertheless useful because it isolates a conceptual issue that is almost purely philosophical from those that are more properly regarded as foundational issues specific to quantum theory; this more purely philosophical issue is the problem of hidden variables, and it forms the focus of this paper.

But how, it might be asked, can progress be attained on *any* of the conceptual issues associated with quantum mechanics without addressing the measurement problem? For a study which, like this one, is focused on the problem of hidden variables, the answer is two-fold. First, the measurement problem comes heavily loaded with assumptions that are simply irrelevant to the statement of the problem of hidden variables. Secondly, insistence on the primacy of the measurement problem illegitimately conflates the concept of the evidential base for the theory with the concept of its intended domain of application. An adequate discussion of the first point is outside the scope of this paper. But there is something that can usefully be said about the second point, some of which bears on the first.

Although quantum mechanics is a theory of the fundamental constituents of matter, it—like any other empirical theory—has an evidential basis which is epistemically distinguished as accessible to us independently of our commitment to its preferred theoretical description. This is evident from the fact that the evidentiary basis was known prior to the discovery and elaboration of the quantum theory. Nor is it contradicted by the fact that the description of the evidentiary base has benefited from the understanding of it that the development of the theory has made possible. This entirely familiar lesson is one that can be told regarding any empirical theory, and it points to a further respect in which the problem of hidden variables has a degree of "purity" not shared by the conceptual issues measurement raises. The problem of hidden variables can be formulated in the language of the evidentiary basis for the theory without invoking the correctness of the quantum theoretical account of the measuring instrument. To concede this leaves entirely open the possibility that although the *problem* is so formulable, its preferred *solution* may not be. But the essential point is that the legitimacy of separating the evidentiary basis of the theory from the theory itself is sufficient to allow for the conceptual possibility that the problem of hidden variables has a formulation and solution that does not depend on the quantum mechanics of measurement. This is the only concession I require for my proposals to warrant consideration.

The paper is organized as follows. I begin with a reconstruction of the quantum logical interpretation advanced in Bub's first book, and in a number of my early papers. Although much of what I have to say—especially in the next two sections—played a major role in the evolution of our thinking, it was never as explicitly addressed in these early writings as I now believe it should have been. I then turn to the discussion of the problem of hidden variables and the bearing of the concept of an effect on its solution.

*1. The status of logic in quantum mechanics*

Priority for the idea that quantum mechanics may have consequences for logic clearly belongs to von Neumann who first noted the possibility in his treatise (1932/1955), and later developed the idea somewhat differently and in considerably more detail in his work with Birkhoff (1936). Within the philosophy of science community, there was the totally different—and almost totally neglected—proposal of Reichenbach in his book of (1944). Despite the existence of such well-known precedents, Quine (1951) is almost invariably the author to whom philosophers turn for the thesis that logic is revisable on the basis of empirical considerations derived from quantum mechanics. Yet the position defended in "Two dogmas" is a highly general methodological one; we may call it

> *Quine's observation.* In case the conclusion of a derivation is contrary to what experience suggests, there is no methodological principle which compels us to exclude the possibility of revising the logical component which figured in the derivation.

Whatever our view of the correctness of Quine's observation, we should be struck by the fact that even if it is correct, it tells us little about logic, and still less about the role of logic in physics.[1] To see this, notice that a conventionalist conception of logic can readily admit the observation, and that Poincaré made the same claim regarding the geometrical component of any derivation of physics while maintaining the conventionality of geometry. Thus Quine's observation certainly does not require that logic should have anything approximating to a *descriptive* content.

Considerations like the preceding might suggest that we should take Quine's observation to be that factual and non-factual components of our conceptual framework are not separable on the basis of their revisability in light of experience. Then the fact that even such a pre-eminently non-factual component as logic is susceptible to revision falls naturally into place: it undermines the notion that the factual and non-factual components of our framework might be distinguished on the basis that the former component is revisable for empirical reasons while the latter is not. On this view, the correct interpretation of Quine's remarks is that it is not even a *part* of his observation that logic is descriptive of anything, since his goal is to undermine the clarity of the distinction between the factual and the non-factual components of our total theory.

But the comparison with conventionalism is also instructive in connection with this view of Quine's observation: unless the logical component can be shown to be revisable because it is presumed *false*, there is nothing particularly striking about the claim that logic might be altered in the face of recalcitrant experience. Here too it is useful to consider the comparison with Poincaré and the example of geometry. For Poincaré a revision of the geometrical component of our total theory would merely reveal a preference for a different set of conventions. But Poincaré's view is tendentious and philosophically interesting precisely because it undermines our pre-analytic conception of the *descriptive* character of geometry. By contrast our pre-analytic conception of logic is

---

[1] Quine appears not to have read the paper of Birkhoff and von Neumann; this may explain his persistent conflation of their work with that of Reichenbach.

*not* one according to which logic is descriptive of reality. This being so, its revision is, in an important sense, subject to *fewer* constraints, not *more*, and so its revisability should be more readily tolerated than the revision of other components of our conceptual framework. Indeed, given the conventionalist climate of opinion at the time Quine wrote, the critical reception of his observation is puzzling. If the logical component is merely a set of conventions, why should it be surprising to be told that it is revisable—even revisable in light of experience? If Quine's observation is to count as a challenge to our view of logic comparable to Poincaré's challenge regarding our view of geometry, it needs to be supplemented. The problem that Quine should have addressed—but did not—is the precise opposite of the one considered by Poincaré in the context of geometry: Quine should have explained how the logical component of our total theory could conceivably be descriptive.

It would make a nice package to be able to say that in his well-known paper, "Is logic empirical?," Putnam filled this lacuna. That the paper begins with a discussion of physical geometry certainly suggests such a characterization of its development of Quine's observation. The suggestion is further strengthened by the fact that, for Putnam, the central epistemological issue is how a revision of classical logic for empirical reasons can be methodologically justified without embracing a particular form of conventionalism about logic.

Putnam argues that the status of logic in quantum mechanics can be usefully compared with the status of physical geometry in general relativity. The point of his comparison is to show that neither the claim that space is non-Euclidean[2] nor the claim that logic is non-Boolean is adequately represented as a change in the meaning of the primitive terms of geometry or logic. The criteria for sameness of meaning in the geometrical case concern the naturalness of the mathematical generalization of a geometrical notion such as that of *straight line* in synthetic geometry by that of *geodesic* in differential geometry. The argument that the geometrical generalization is natural is straightforward, and Putnam presents it elegantly and well. He then argues that the case of logic can be given a similar treatment by observing the considerable overlap in the formal properties which are shared by the quantum logical analogs of the classical connectives *or*, *and* and *not*. But if this were all Putnam had shown, his account would not have addressed the sense in which logical laws might be said to be descriptive. At most he would have shown that the laws of "quantum logic" are also laws involving the familiar logical connectives. It would remain open for a conventionalist about logic to maintain that the laws that the logical connectives are held to satisfy are accepted merely for the convenience they effect in the statement of our total theory. Such a conventionalist could even accept Putnam's argument that there has been no change of meaning involving primitive terms and argue that just as Poincaré showed that geometry requires a *decision* about which objects to count as straight lines, so also, logic requires a decision about which propositions to count as tautological.

---

[2] Putnam is aware that this is an over-simplification.

There are however two respects in which Putnam might reasonably claim to have significantly extended Quine, neither of which is contained in his discussion of the change of meaning issue. The *first* of these extensions revolves around Putnam's exposition of the methodological basis for changing logic. Putnam claims that by changing the logical component of our total theory it is possible to address the outstanding *conceptual* difficulties of quantum mechanics within a *realist*, but *non-hidden variable*, framework. This has its correlate in the avoidance of what, following Reichenbach (1958), Putnam calls "causal anomalies." Had Putnam established this claim, he would have gone considerably further than Quine's allusion to recalcitrant experience; moreover, it is conceivable that this claim is capable of being justified without directly addressing the question whether logic plays a descriptive role in fundamental physics. The *second* respect in which Putnam attempts to extend Quine's observation *does* address logic's descriptive content. This is Putnam's discussion of the "physical interpretation" of the logical connectives as operations on "tests," where, as Putnam acknowledges, his account borrows from earlier work of Finkelstein (1962).

At this point, I would like to briefly consider the connection between Putnam's logic of tests and his claim to have provided a realist non-hidden variable solution to the conceptual problems of quantum mechanics. The context of my discussion is Putnam's early execution of his quantum logical interpretation. Later I will review Putnam's mature quantum logical proposal for addressing the conceptual difficulties of the theory.

Putnam argues that the lattice of closed linear subspaces of Hilbert space is interpretable as a lattice of tests for the dynamical properties of a quantum mechanical system. It is an explicit part of his view that the proposed interpretation supports counterfactual reasoning about the properties associated with such tests:

> Let us pretend that to every physical property $P$ there corresponds a test $T$ such that something has $P$ just in case it 'passes' $T$ (i.e. it *would* pass $T$, if $T$ were performed).[3]

Hence, even in the absence of carrying out a test for $P$ we may consider the hypothetical possibility that a system $S$ has $P$ since this is equivalent to the counterfactual that if $T$ were performed $S$ would pass $T$. More precisely, suppose $P$ and $Q$ are possible properties of $S$, and that $T_P$ and $T_Q$ are ideal measurement procedures which test whether $S$ has the properties $P$ and $Q$ (respectively). Suppose also that there is no measurement procedure which simultaneously tests for both $P$ and $Q$. Finally, suppose that $T_Q$ is performed and a result found. Then according to Putnam, although $T_P$ was not performed, it is nevertheless the case that $P$ holds of $S$ or does not hold of $S$, and one or the other alternative would have been found had the test $T_P$ been carried out instead of the test $T_Q$.

Now the natural interpretation of this proposal is that it is implicitly a hidden variable interpretation of quantum mechanical propositions (of which the properties $P$ and $Q$ are constituents) in terms of a space of possible truth value assignments. These truth value assignments may be epistemically inaccessible to us, but that there are such assignments appears to be essential to the proposal. In any event, nothing in Putnam's paper explicitly precludes such an interpretation, and as we have seen some of his remarks implicitly assume it. But then the claim to have offered a realist interpretation that is not a hidden variable interpretation has to be given up. Although Putnam later came to elaborate an alternative interpretation of his remarks, one that is not obviously excluded by foundational developments which occurred contemporaneously with the writing of his paper, the alternative is also problematic. Precisely how it is problematic is a point to which I will recur.

---

[3] Putnam (1968; p. 195 of its reprinting in his 1975); the emphasis is Putnam's.

But how, it might well be asked, can Putnam be said to have committed himself to a hidden variable interpretation when one of the stated aims of his paper is the avoidance of hidden variables? This is a reasonable question, and it deserves a much more extended answer than I can give it here. But in essence the answer is that the notion of a hidden variable interpretation that I have appealed to in my criticism of Putnam is not the notion Putnam intends when he claims to have achieved a realist, non-hidden variable interpretation of quantum mechanics. By claiming to avoid hidden variables, Putnam almost certainly meant that he could avoid a theory like Bohm's or de Broglie's. For Putnam, such theories are characterized by their invocation of special forces to account for the disturbance by measurement.[4] Since he nowhere appeals to such forces, Putnam believes his account to be free of an appeal to hidden variables. The idea that his interpretation of the logical connectives assumes an underlying space of truth-value assignments—and that it is therefore a hidden variable interpretation of quantum mechanics after all—appeals to an abstract notion of hidden variable theory, one which was only beginning to emerge when Putnam wrote his paper, and one with which he was at the time almost certainly unfamiliar.[5]

This is perhaps a good place to pause and consider for a moment the point to which we have gotten. A common conception of Putnam's quantum logical interpretation is that it was inspired by Quine, and that it was the fulfillment of Quine's prophetic observation regarding the revisability of logic. But if I am right, this view must be understood in the context of the following facts: Quine's observation has almost nothing to say that is of specific relevance to logic. It is easy to argue on general grounds that logic is revisable. The difficult argument to make out is the specific one, that by revising logic it is possible to interpret the theory realistically and without hidden variables. Putnam's attempt to develop such an argument and show that Quine's position has interesting implications for logic succeeds only in relapsing into a kind of hidden variable interpretation of the theory. As a result, the methodological promise of Putnam's account—that it would address the excessive generality of Quine's observation—simply evaporates. Finally, the fact that both Quine and Putnam oppose a fact/convention distinction seems to have limited, rather than enhanced, their ability to formulate an interesting claim about logic and its relation to physics. These early explorations of the status of logic in quantum mechanics are mainly of interest for showing how little progress was possible without greater familiarity with the new developments in axiomatics, and the revival of the problem of hidden variables which these developments fostered. In this respect, the epistemological situation is interestingly similar to the case of non-Euclidean geometry before the discovery of the relative consistency proofs; in both cases the philosophically most interesting discovery was a meta-theoretical one. Non-Euclidean geometry remained a mere curiosity, one with little philosophical relevance, until it was discovered that it could be modeled within Euclidean geometry. This discovery, especially in the form of Poincaré's beautiful ruler and compass constructions, undermined the intuitive authority of Euclidean geometry, and explains why Poincaré occupies such a central position in the philosophy of the subject—whether or not one follows him in his conventionalism. In the case of logic, the principal discovery for epistemology is the *absence* of an inner model of its classically divergent logical structures in a Boolean algebra.

---

[4] I am not endorsing this characterization; I am merely reporting what I believe to have been Putnam's view of hidden variable interpretations and the theories on which they might be based.
[5] Pitowsky (1989, §4.5) critically discusses the reconstruction of Putnam's remarks in terms of an underlying space of non-standard truth value assignments..

2. *The possibility structure of physical systems*

I want to set aside until later further discussion of the quantum logical interpretation of quantum mechanics, while I briefly consider the contribution of axiomatic quantum logic to the formulation of a framework within which to conceptualize the theory's novel features and its particular challenges. For the issues I will be addressing this framework is notable for the centrality it accords the generalization of the notion of a Boolean algebra and of a probability measure. As such, its connection with logic is at least two-fold: the algebras it considers are algebras of (physical) *propositions*, and the metatheory of ordinary logic is closely related to the representation theory of Boolean algebras. The principal significance of this framework for our earlier discussion of the status of logic in quantum mechanics is the suggestion that it is the algebraic structure of the propositions to which probabilities are assigned that forms the proper subject matter of an empirically mandated change in logic. To propose a change in logic on the basis of developments in physics is to move from a Boolean algebra of propositions to one of the generalizations associated with the partial Boolean algebra or orthomodular poset of subspaces of a Hilbert space. Such a transition in the algebra of *propositions* is comparable to the transition from a Newtonian to a Minkowskian geometry of *events*. Boolean algebras and their quantum mechanical generalizations constitute assumptions about the *possibility structure* of propositions, just as Newtonian and Minkowskian geometry constitute assumptions about the *spatio-temporal structure* of events.

On the assumption that some set of principles relating to the possibility structure of propositions is unavoidable, the interest of the proposal that quantum mechanics involves such a shift in conception is that it might yield an explanation of the special class of probability measures one finds in the theory. Indeed, a natural interpretation of Gleason's (1957) theorem—that the generalized probability measures definable on a Hilbert space are precisely those given by the Born rule—is that it explains the quantum mechanical probability assignments by appealing to the algebraic structure of physical propositions. This idea is of fundamental importance in both of Bub's books, and it continues to exert a powerful influence on his thinking about these matters. It is also in one sense interpretationally neutral, since it does not depend on a quantum logical interpretation of the theory. However, there is another sense in which it is not only *not* interpretationally neutral, but stands in the way of achieving clarity about the problem of hidden variables. But before addressing this claim, it will be worthwhile to have before us a statement of the central no hidden variable theorems. Among other virtues, the formulation given in the next section yields an elegant comparison of the Kochen-Specker theorem with Bell's no hidden variable theorem, and this will allow me to clarify my reasons for focusing on the theorem of Kochen and Specker.

*3. The no hidden variable theorems*

The formulation which follows is due to Pitowsky who presented it to me in correspondence.

Let $F$ be a finite set $\{A, B, C, \ldots\}$ of Hermitian operators acting on a finite dimensional Hilbert space $\mathbf{H}$; the operators in $F$ represent a family of dynamical variables with possible values corresponding to the eigenvalues of the operators. A quantum state $\rho$ assigns a probability distribution $P_\rho$ to the possible values of the dynamical variables represented by the operators in $F$. For each possible value $a_1, a_2, \ldots$ of $A$ we have $P_\rho(a_1|A), P_\rho(a_2|A), \ldots$; for each value $b_1, b_2, \ldots$ of $B$ we have $P_\rho(b_1|B), P_\rho(b_2|B), \ldots$, etc. (The set $F$ may be so chosen that the probabilities associated with this finite set completely characterize $\rho$ as a state on $\mathbf{H}$, but we make no use of this fact here.)

As observed by Kochen and Specker, there exists a *classical* probability measure (the product measure),

$$P_\rho(a, b, c, \ldots|A, B, C, \ldots) = P_\rho(a|A) \, P_\rho(b|B) \, P_\rho(c|C)\ldots,$$

on the operators in $F$ provided we ignore algebraic relations among them and assume that the dynamical variables they represent are all independent of one another.

Suppose, however, that we require of $P_\rho$ that it satisfy the following two conditions:

(1) $P_\rho(a, b, c, \ldots|A, B, C, \ldots)$ is a probability measure defined on all eigenvalues of all the operators $A, B, C, \ldots$ in $F$;

(2) If $A, A', A'', \ldots$ in $F$ *commute*, then $P_\rho(a, a', a'', \ldots|A, A', A'', \ldots)$ coincides with the quantum mechanical probability assigned by $\rho$.

Then *Bell's no hidden variable theorem* tells us that there is an $F$ and a $\rho$ such that $P_\rho$ does not exist. (In this case the operators in $F$ are local operators acting on a tensor product of Hilbert spaces.) And *Kochen and Specker's no hidden variable theorem* tells us that there is an $F$ such that for any $\rho$, the distribution $P_\rho$ does not exist.

Pitowsky (1994) shows that the existence of $P_\rho$ is equivalent to the requirement that the numbers

$$\{P_\rho(a, a', a'', \ldots|A, A', A'', \ldots): A, A', A'', \ldots \text{ in } F \text{ commute}\}$$

satisfy a finite family of linear inequalities, which Boole called "Conditions of possible experience." The non-existence of $P_\rho$ implies the failure of at least one such inequality. But what is of most importance for our purposes is the fact that if $P_\rho$ exists, it is representable as a weighted sum of two-valued measures on commuting $A, A', A'', \ldots$ in **F**. The incompatibility of the probabilities of quantum mechanics with the classical use of two-valued measures is therefore an implication that *both* theorems share.

There are several reasons why I have chosen to focus on Kochen-Specker, rather than Bell. First, there is an obvious sense in which the fact that there are no two-valued measures is more fundamental than the fact that quantum mechanical probabilities are not representable as the weighted sum of such measures—even when they do exist. Secondly, as Pitowsky's formulation makes evident, since the theorem of Kochen and Specker is independent of the choice of quantum mechanical state $\rho$, it uses less of the theoretical apparatus of the quantum theory for is formulation. And finally, if what I have to say regarding the significance of the Kochen-Specker theorem is cogent, it should extend to the case of Bell's theorem as well.

It is well-known that to formulate the essential physical content of the Kochen-Specker theorem we can dispense with the notion of a finite family of Hermitian operators acting on a finite dimensional Hilbert space and consider instead the finite family of properties associated with the possible point values of the dynamical variable each such operator represents. The association of such properties with propositions about a physical system and that of a probability measure over such propositions is an established part of classical probability theory. The mathematical generalizations of algebras of propositions and of probability measures over them, though motivated by the case of quantum mechanics, can be expressed with very little technical machinery and without special reference to the quantum theory. Thus, even though the theorem of Kochen and Specker might never have suggested itself were it not for the development of quantum mechanics and the presumption that it might bear on the foundational problems mentioned earlier, the conceptual issues the theorem poses are rightly distinguished as philosophical rather than foundational because the notions on which their formulation depends are of such generality that they are held in common by virtually every theory.

By the coincidence of two-valued probability measures with truth value assignments, the notions with which the problem of hidden variables is concerned are precisely those of *property, proposition, probability* and *truth*. This is why the problem is of such generality and why, among the conceptual issues raised by quantum mechanics, it is the most purely philosophical. Within this abstract setting, the question I wish to consider is how best to understand the significance of such configurations of propositions, configurations that do not admit two-valued generalized probability measures. What does their possibility tell us about the conceptual framework of a generalized probability theory that allows for the possibility of such configurations?

In the case of interest to us—namely quantum mechanics—the probabilities concern the behavior of elementary particles and atoms, as well as simple composite systems comprised of combinations of them. For the purposes of understanding the significance of the no hidden variable theorems, it suffices to consider composite systems of at most two or three particles; indeed systems of two particles suffice to generate the issues surrounding locality that are raised by Bell's theorem. Although I will generally focus on single particle systems, the discussion is intended to apply to such composite systems as well.

I will assume that particles are epistemically accessible to us by virtue of their association with the variety of observable *effects* they produce, effects whose production we may examine at will in a variety of controlled settings. Effects are typically understood as indicators of the properties of particles, but this is an assumption we will find reason to question.

I will say that a *proposition belongs to a particle* if its constituent property is a possible property of the particle. The standard interpretation of a generalized probability theory of the sort found in quantum mechanics is that its probabilities are probabilities of propositions belonging to particles. The idea I propose to develop is that quantum probabilities are not satisfactorily interpretable as probabilities of propositions belonging to particles. I will argue that such an interpretation is defensible only when the propositions belonging to a particle form a Boolean algebra (or an algebraic structure that is at least homomorphic to a Boolean algebra, a qualification I will henceforth assume, but not mention). I will then outline an alternative framework, a framework according to which the probabilities are probabilities of effects rather than propositions. But before elaborating this idea, it is important to distinguish it from another similarly expressed proposal with which it should not be conflated.

*4. The propositional framework and the failure of determinacy*

A widespread view of the significance of the problem of hidden variables is that there are dynamical properties which, although possible properties of a particle, are properties it sometimes neither has nor lacks. On this view, the import of the no hidden variable theorems is not a mere failure of determinism, but a failure of determinacy. By *determinacy* I mean the general thesis regarding properties according to which nothing can have a generic property without having one of the specific properties that fall under it. For example, nothing can be colored without being some specific color, nothing can have a shape without having some specific shape, etc. The thesis of determinacy is interchangeable with one about the truth of propositions, since a proposition which ascribes a determinable property is necessarily equivalent to the disjunction of all those propositions that ascribe a determinate property which falls under the determinable. Thus, when formulated in terms of the truth of propositions rather than the holding of properties, determinacy is the thesis that such a disjunction is true only if exactly one of its disjuncts is true.

The intuition that supports determinacy is that while it makes perfect sense, when thinking of a fictional world, to treat some propositions belonging to its inhabitants as neither true nor false, it remains to be explained how this idea should be understood when we are concerned not with fiction, but with reality. This is perhaps one of the distinguishing marks of a fictional world. It may be possible to explain how the denial of determinacy should be understood outside the context of fiction, but so far as I am aware, no one has taken on this task. Allusions to potentiality and related notions have never risen above the rank of mere allusions.

In discussions of quantum mechanics the propositions that are thought not to be determinately true or false are those whose constituent properties are associated with the point values of a dynamical variable. For example, the denial of determinacy is a basic tenet of all modal interpretations, and of many others besides. It is usually justified by a consideration like the following: A determinable dynamical property may hold even though none of its determinate properties holds, since having the determinable property is necessarily equivalent to the truth of the disjunction of the propositions whose constituent properties exhaust the possible point values of the dynamical variable. Such a disjunction is always true, but by Kochen-Specker it is not possible that for every such disjunction, exactly one of its disjuncts should be true and the others false. Hence the lesson of their no hidden variable theorem is indeterminacy.

There is considerable latitude in how determinacy has been addressed in the philosophical discussions surrounding quantum mechanics. However, *all* approaches assume what I will call a *propositional framework*; that is they all share the idea that probabilities are defined over propositions which *belong to* a particle. This is true both of solutions to the problem of hidden variables which respect the algebraic structure of the propositions associated with Hilbert space, and of those which do not. Let us call the former solutions *algebraic solutions*, and the latter, *hidden variable solutions*. Algebraic solutions operate under the assumption that the structure of physical propositions is a real constraint, one that is as significant for our understanding of the problem of hidden variables as the structure of space-time is for our understanding of the constancy of the velocity of light.

With one exception (to be discussed in a moment) it is the algebraic solutions that accept some degree of indeterminacy. The key difference among algebraic solutions consists, in the first instance, in how extensive the class of propositions which are determinately true or false is taken to be, and in the second instance, in how the selection of the class of determinate propositions is justified. For example an algebraic solution to the problem of hidden variables might make the selection of the class of determinate propositions depend on a connection between the probability 1 or 0 assignments of the quantum wave function and truth and falsity; or, as on Bub's modal interpretation, it might be more subtle than this. But however the selection is effected, the class of propositions that are determinately true or false must be extensive enough to explain the fact that measurement always yields a determinate property; at the same time, it must not be so extensive that it precludes the existence of a two valued generalized probability measure on the selected class of determinately true or false propositions belonging to a particle. This means that for an algebraic solution there will almost always be some indeterminacy.

The algebraic solution that attempts to preserve the determinacy of all propositions is the quantum logical interpretation. Here I refer not to Putnam's first formulation of the idea, which we considered earlier (although it too seeks to uphold determinacy), but to later formulations, such as those in Friedman and Putnam (1978), Bub (1974) and Demopoulos (1977). This approach promises to deliver a determinate truth value for *every* proposition by appealing to the consideration that a complete disjunction of determinate propositions under a common determinable is true only if exactly one of its disjuncts is true; it takes this condition to be satisfied because such a disjunction coincides with the unit of the algebra of propositions, and the unit proposition is always true. The non-classicalness of the logic of propositions consists in the fact that while all such disjunctions are true, it is not possible to represent this fact by an assignment of truth values, as it is in the case of classical logic.

There are various objections to the quantum logical interpretation, but in the case of its defense of determinacy, the central difficulty is that the notion of truth to which the interpretation appeals, is one that validates classical contradictions. To see this, recall that the finite partial Boolean subalgebra of subspaces of $\mathbf{H}^3$ isolated by Kochen and Specker can be represented by a propositional formula (Boolean polynomial) in 86 variables. For every assignment of elements of the two element Boolean algebra to its variables, the formula takes the value 0, and is therefore a classical contradiction. But there is an assignment of subspaces, representing quantum propositions, under which it is associated with the unit of the partial Boolean algebra of subspaces of $\mathbf{H}^3$, and is therefore satisfiable. Hence, if we think of the laws of logic as "laws of truth"—as a guide to our concept of truth—the quantum logical interpretation's concept of truth cannot be our classical concept. In the only other foundationally important case of a proposal of an alternative logic—namely, Intuitionism in the foundations of mathematics—the proposed alternative is *conservative* over classical logic in the sense that neither classical contradictions nor the denials of classical tautologies are validated. Moreover there is, in the Intuitionist case, an intelligible notion of truth on which its logic rests, that of constructive provability. This notion has a content which is capable of being understood independently of the proposal to change the logic which underlies mathematical reasoning and metalogical studies; moreover, the explanation of the Intuitionist notion of truth in terms of constructive provability enables us to understand why, for example, *P* v ¬*P* fails in Intuitionist logic—even if we are convinced of the correctness of classical logic. But the quantum logical concept of truth—coincidence with the unit of the algebra of propositions—is a purely formal one. And in view of the fact that this notion of truth allows for the satisfiability of classical contradictions, the interpretation rests on a notion of truth that it cannot simply take for granted, but which it also appears unable to elucidate.

Let me turn now to hidden variable solutions to the problem of hidden variables. There are various ways of implementing a hidden variable defense of determinacy, but the one that appears to have the most natural physical motivation is based on the idea that properties and propositions must be individuated more finely than the algebraic structure of Hilbert space suggests. The methodological basis for the implementation of this idea is that every property requires a clear physical criterion for saying when it holds and when it fails to hold. This is the principle of Einstein which Heisenberg sought to emulate—the "good joke that should not be repeated too often," as Einstein is supposed to have said of it. To make the proposal tractable, we may imagine a special case of dynamical properties, one for which the class of operational procedures is specified in some simple way; for example, the procedures might be indexed by triples of orthogonal directions in physical space, with the directions corresponding to orientations of the experimental apparatus. Without further information, the measured properties may be assumed to be so finely distinguished that they have no operational procedures in common. An elementary physical proposition is then defined by its constituent property and *the* operational procedure which is criterial for it and for its negation. Under this scenario, there may be relations of statistical equivalence among elementary propositions, but the propositions themselves are distinct, a fact which is reflected by the differences among measurement procedures which are criterial for them. Such propositions may differ in truth value, even though they are statistically equivalent.

There is a general methodological fault with such an approach, one which arises in connection with the *objectivity* of its concept of a property. The relevant notion of objectivity is one according to which a property is conceptually independent of the canonical means of accessing it: the same property should be accessible in different measurement contexts, and by alternative measurement procedures. By tying the notion of a property to a single operational procedure, we renounce this aspect of our concept of objectivity, since we thereby abandon the idea that the *same* property may be presented differently. The difficulties which derive from the relativity to measurement context in the case of composite systems serve only to place this fact in an especially problematic light. Insofar as hidden variable approaches to determinacy have a fundamentally realist motivation—and determinacy is certainly a consideration which derives from realism—their fundamental notions appear to be in tension with one another: relativity to the measurement context secures the reality of dynamical properties—which is to say their determinacy—at the price of their objectivity. To concede that the same property may have been presented differently—i.e., as the property indicated by a number of different measurement procedures—is tantamount to admitting counterfactual reasoning regarding the propositions belonging to a particle. But counterfactual reasoning in association with a multiplicity of measurement procedures is indispensable to the objectivity of the constituent properties of the propositional framework. Hence, a great deal of the counterfactual reasoning which supports our conception of physical properties as objective is rendered vacuous when properties are bi-uniquely paired with operational procedures.

Although it is generally conceded that the Kochen-Specker theorem is mathematically more significant than von Neumann's treatment of the problem, it is not uncommon to see their theorem dismissed as having no greater *interpretational* interest than von Neumann's on the ground that their argument can be circumvented if properties are relativized to a measurement context. But this misses the philosophical advance of their analysis. By contrast with the theorem of von Neumann and others, the analysis discloses the connection between determinacy, contextuality and objectivity in a realist propositional framework. This is the principal implication of the fact that the theorem is proved under the restriction to a *partial* algebra of propositions. We have just seen that non-contextuality is as fundamental to a realist interpretation of the theory as determinacy, since it is precisely when properties are made relative to a measurement context that counterfactual reasoning about them is rendered vacuous, and their objectivity compromised.

The necessity of having to choose between objectivity and determinacy does not pose a dilemma for the Copenhagen Interpretation. It also accepts the primacy of the propositional framework, but since it does not even purport to be a classical realist interpretation of the theory, it is not committed to the objectivity of the properties of particles. It can therefore accept the relativity of a property to a measurement context as confirmation of its view that quantum reality, unlike classical reality, does not support the notion of a "detached observer." But there are also non-dynamical or "eternal" properties of particles. The difficulties which afflict the probabilities of dynamical properties do not arise for such eternal properties, and a detached observer concept of reality appears to be fully justified for *them*. This suggests that our focus on dynamical properties and the concept of reality they support may be a poor guide, and that we should look elsewhere than the propositional framework if we are to have a viable conception of the peculiar probability assignments which the no hidden variable theorems reveal.

*5. A framework of effects*

The formulation of the quantum theory in terms of the propositional framework of axiomatic quantum logic was essential to understanding the status of probability in the theory, since it was only after this step was taken that it became clear that there are difficulties with interpreting the theory's probability assignments that are independent of the measurement problem and the paradoxes. This is another respect in which early formulations of the quantum logical interpretation obscured the foundational situation. Instead of providing a solution to the long-standing issues involving measurement and the paradoxes, the discovery of "the logic of quantum mechanics" revealed a new and different problem, namely, the impossibility of interpreting the probabilities of the theory so that every proposition belonging to a particle is non-contextually determinately true or false. Rather than refuting classical logic, this discovery should be seen as refuting the idea that the probabilities are interpretable as the probabilities of such propositions. Ironically, the consequences of quantum mechanics for logic are almost the precise opposite of what Quine and Putnam imagined they might be. If anything, the problem of hidden variables upholds the centrality of classical logic in our theorizing about the physical world, while allowing that the Boolean algebraic structures, which are so closely associated with classical logic, are not appropriate for every use of probability in physics. Thus, the essential idea of the present proposal is that the no hidden variable theorems possess a kind of "dialectical" significance—much in the sense of DiSalle's (2006) execution of this idea in his study of the evolution of theories of space-time—showing the necessity of replacing the propositional framework on which they are based by another altogether different framework, and, in the process, rejecting both algebraic and hidden variable solutions to the problem of hidden variables.

The probabilities of "quantum probability theory" are not defined over the totality—or even a sub-totality—of the propositions belonging to a particle, but over the totality of *effects* which are induced by an interaction with an experimental set-up and registered in the experimental apparatus. Since effects comprise the accessible historical record of such interactions, the issues we canvassed earlier surrounding the objectivity of properties and propositions do not arise for them. Unlike the constituent properties of propositions belonging to a particle, effects are not merely indicated by the experimental arrangements which comprise measurements; hence the reality or objectivity of an effect is not in any way tied to the legitimacy of counterfactual reasoning regarding measurements of properties of particles which might have been performed, but in fact were not.

From the point of view of the effects framework, the logical form of the representation of an elementary particle is not that of a class of propositions whose constituent properties are possible properties of the particle, but that of a function. By this I mean that aside from its eternal, non-dynamical properties, a particle is characterized by the fact that it is an object which, when subjected to a specified class of operational procedures, produces a family of characteristic effects. A particle is implicated in an essential way in the production of the effects associated with its interactions with experimental set-ups, and of course with the naturally occurring conditions that experiments emulate, and whose influence experiments exhibit in a controlled setting. But the representation of a particle as a function is not committed to the idea that a particle would have had some dynamical property had it been presented with a different experimental set-up than the one with which it was in fact presented. By contrast, the representation of a particle by a class of propositions is committed to the truth of a great many such counterfactual claims about its properties.

The representation of an elementary particle as a function which, when presented with an experimental configuration, yields an effect, is interchangeable with its representation as a class of propositions only when the effects are predictable with 0-1 probability. The fact that the particle-effects described by quantum probability theory are not so interchangeable with propositions belonging to particles may or may not figure interestingly in the solution of some of the conceptual problems associated with the theory that were mentioned earlier. But it clearly bears on the problem of hidden variables since, although there is no deterministic theory of effects, they are unproblematically determinate.

Within the framework of effects, the problem of determinism can be posed as follows: Given a particle and a class of experimental procedures, to predict the particle-effects with perfect knowledge, i.e. to predict with probability 0 or 1, uniformly and without foreknowledge of the experimental procedure to which the particle will be subjected, the answer to every question regarding the occurrence of a possible effect. The no hidden variable theorems show that the quantum probabilities of such effects are not compatible with the existence of a function which predicts every possible effect with probability 0 or 1; hence, neither is it possible to devise a reconstruction of the state of a particle on which to base such predictions: the probabilities of the effects are logically inconsistent with the existence of such a state.

Within a propositional framework, the closest analogue of a classical state of a system $S$ is a selection of truths from some subset of the Boolean algebras associated with $S$. Since propositions are built up out of their constituent properties, such a state is a catalogue of $S$'s properties. Thus, within the propositional framework, a Hermitian operator $A$ acting on a (finite dimensional) Hilbert space **H** represents a family of possible properties of $S$ (or a family of propositions which belong to it). Another Hermitian operator $B$, whose properties are included in those represented by $A$, is given by some function $f$ of $A$. But $B$ may also be given as a function $g$ of another operator $C$. If we apply a measurement procedure $M_A$ for $A$ to $S$ and obtain the result $a$, it seems natural to infer that $S$ has the property $b = f(a)$; similarly it seems natural to suppose that applying $M_C$ to $S$ would yield the result $b = g(c)$, and so on for all the operators $A, B, C, ….$

The framework of effects differs from the propositional framework by treating the Hermitian operators $A, B, C, …$ as representative of operational procedures $M_A, M_B, M_C, …$ associated with $A, B, C, …,$ rather than representative of families of properties or propositions.[6] Applying $M_A$ to $S$ does not yield a proposition belonging to $S$ as its result, but an effect that is recorded in the apparatus associated with $M_A$. Because $A$ and $B$ are functionally related we can calculate that had $M_B$ been performed rather than $M_A$ the result would have been $b = f(a)$. Similarly, since $B$ is functionally related to $C$ by $g$, if $M_C$ had been performed instead, the result of performing $M_B$ can be calculated on the basis of the functional relation $g$ that holds between $B$ and $C$: the result of performing $M_B$ would have been the outcome $g(c)$.

Now in the case of properties, it is natural to think that properties of $S$ are (1) indicated by a multiplicity of measurement procedures, but (2) hold independently of the application of any of these procedures, so that (3) whether or not all the procedures $M_A, M_B, M_C$ are performed, $S$ has properties associated with each of $A, B, C$, and these properties reflect the functional relations among $A, B, C$. This leads to well-known difficulties when the families of possible properties of $S$ are interrelated after the manner of their quantum mechanical representation. That effects are *not* constrained by either assumption (1) or (2) is a simple consequence of the concept of an effect, one that follows without the supplementation of anything corresponding to a "metaphysics of effects." This contrasts with the case of properties, where both the acceptance and the rejection of assumptions (1) and (2) lead very quickly to controversial philosophical theses. The failure of (3), which is revealed by the no hidden variable theorems, mandates some modification of the underlying concept of a property within the propositional framework. But so far as effects are concerned, the fact that we cannot predict the effects that will be produced by $M_A$, $M_B$, $M_C$, … is merely a manifestation of indeterminism, i.e., of the impossibility of complete knowledge of $S$'s behavior under the totality of circumstances $M_A$, $M_B$, $M_C$, ….

---

[6] It was observed by Hultgren and Shimony (1977) that the spin propositions of a spin-1 system do not exhaust the partial Boolean algebra $B(\mathbf{H}^3)$ of subspaces of a three dimensional Hilbert space $\mathbf{H}^3$. This is in contrast with spin-1/2 propositions and $B(\mathbf{H}^2)$—as Shimony and Hultgren also observed. The context in which Shimony and Hultgren made their observation is relevant to the proposal in the text, which assumes that every Hermitian operator is represented by an operational procedure; they raise the question "Is it possible to give an operational motivation for the whole of $B(\mathbf{H}^3)$"? Their question was answered positively by (Reck et al. 1994). I am grateful to Itamar Pitowsky for drawing my attention to these two papers.

What is given up in the transition from classical probability to quantum probability is not the determinacy of effects, but the notion that the quantum theory in any way depends on the existence of a "catalogue" of propositions belonging to a particle—a catalogue of those which have been found to be true and those which would have been found true, had the particle been subjected to a different experimental procedure. The effects framework requires something much less, namely, the existence a catalogue of those effects which have been found, and those that might have been found had the particle been subjected to other experimental set-ups. This idea is entirely in conformity with the view that Fuchs and Peres (2000a and 2000b) express with the slogan, "Unperformed measurements have no results." Let us review some of the difficulties which conformity with this slogan has been thought to face.

*The effects framework is irremediably instrumentalist.*

I take this to mean that the transition to effects represents a fundamental physical theory as a mere heuristic for prediction, with the consequence that "… physics could only claim the interest of shopkeepers and engineers."[7] While instrumentalism may be a consideration that has attracted some proponents of the "framework of effects," it is by no means essential to it. What is at issue is the interpretation of the quantum algorithm for assigning probabilities and for reasoning from the position of uncertainty which such probabilities necessarily signify. This leaves a considerable degree of realism intact. First, there is nothing in the view that denies the existence of elementary particles; nor does the view seek to represent a particle as a logical construction out of its effects. Secondly, the view allows that there is a plethora of properties which elementary particles are unproblematically represented as having; these include all of their eternal or non-dynamical properties. Both they and the particles themselves are real objects of theoretical investigation.

---

[7] Einstein; Letter to Schrödinger of December 22,1950 (Przibram, 1986, p. 39).

By replacing the propositional framework with one of particle-effects, we achieve not an ontological economy but a different conceptualization of the significance of the implicit probability theory of quantum mechanics. Since particles are an essential component in their production, the idea behind the proposal is not to eliminate the role of particles in the production of effects but to provide a conceptual framework within which the incompleteness—i.e., the fundamentally probabilistic character—of quantum mechanical predictions may be understood. The point is not to achieve an ontological economy, but to advance a proposal regarding the philosophical novelty of at least one aspect of the theory: the theory represents reality as confronting us with an "inherent incompleteness" in what can be known of particle-effects.[8] Particles are peculiar insofar as the view rejects the usual conception of dynamical properties and their representablity in a propositional framework; but particles and their non-dynamical or eternal properties remain a basic part of the framework of effects. Here is an analogy that might illuminate the bearing on realism of the view of quantum probability that I am advancing.

Our knowledge of the past is constructed from the system of traces it has left in the present. Now imagine that we are concerned to construct a model of past events of such a character that there is very little basis to suppose that they resemble the events with which we are familiar. Suppose also that the traces of these events are accessible to us only in fragments that can be examined one at a time, that the information contained in any one fragment is insufficient to determine a complete account of the events which produced the traces which comprise it, and that the traces are themselves continually in flux. Suppose further that the fragmentary traces do not combine to give a single consistent story describing the truth regarding the events at this earlier time. Now it is conceivable that although the past is in this way "hidden" from us, our epistemic situation with respect to its traces is systematic and even susceptible of a relatively simple representation. Although systematic, the representation of available traces not only fails to facilitate the reconstruction of this past state of the world, but actually precludes the possibility of a consistent reconstruction on its basis. Under such circumstances we might cease looking for a representation of a past state in terms of the propositions belonging to it, because we will have come to recognize that there can be no convergence from present or future traces to such a propositional representation. We might then dispense with the search for a theory of such states and focus instead on understanding the distribution of present traces, their relevance to one another, and the task of predicting their likely evolution. This would be a theory of past events of a sort, but not what we had originally imagined such a theory to be. In particular, it would not aim to *model* the past, but to anticipate its present and future traces. To recover quantum mechanics from the analogy, replace traces with probability distributions over particle-effects, and past states of the world with propositional representations of particles. Then two things are worth noting: neither such a theory of the past nor quantum theory contravenes the thesis of determinacy, and where the one theory accepts the reality of the past, the other accepts the reality of particles. In each case, one has merely abandoned a familiar style of theorizing and modeling.

---

[8] The notion of inherent incompleteness was introduced in my (2006). Although the framework of that paper is a propositional one—a framework which I now reject—inherent incompleteness is easily transferred to the framework of effects.

*The framework has nothing to offer for future research.*

The idea of a theory of effects suggests a general feature of interpretations of quantum mechanics that are expressed within a propositional framework: All such interpretations elaborate a "story" regarding the dynamical properties of a particle, but every such story must address the following dilemma: If the story leads to new and quantum-mechanically divergent predictions, then it is unclear in what sense it is an interpretation; what seems rather to be on offer is an alternative theory. But if such interpretations preserve the probabilistic predictions of quantum mechanics, it is unclear what role the attendant story is playing; after all quantum mechanics achieves the same result without any such story. A brief comparison with one such interpretational story may help to further clarify the present proposal.

Dürr, Goldstein and Zanghi (2003) develop an extended analysis of observables other than position; as we will see, position plays a distinguished role in their conception. The analysis focuses on the case of spin. If I have understood them correctly, when Dürr et al. claim that *spin is not real*, they mean that any association of a random variable $Z_A$ with a self adjoint Hermitian operator *A* representing a spin observable must reflect the fact that the result of a measurement of *A* does not depend on *A* alone, but depends as well on the experimental context. Different random variables will be associated with different measurement contexts; moreover, the different random variables are not constrained to have the same values in different contexts, although the distribution of their values must agree. Hence the association of a random variable with an operator is many-one, involving not $Z_A$, but $Z_{AB}$, $Z_{AC}$, … depending on the experimental context, which determines whether *A* is measured with *B*, or with *C*, or ….[9]

On the surface this appears to be merely a contextualist account of spin properties. But the position is more subtle than this. Dürr et al. hold that since a contextualist account of spin is necessitated by the no hidden variable theorems, the correct conclusion to draw is not that spin properties are contextual, but that because we are required to regard them as contextual, they are really not properties at all. Drawing on Bell's discussion of de Broglie-Bohm, Dürr et al. grant that *something* is discovered in a "measurement of spin," but to understand what this is we have to recognize the priority which attaches to particle position. On their view, spin measurements are in reality measurements of particle positions, and it is particle positions that comprise *the* dynamical properties of particles. Dürr et al. evidently regard it as an advantage of their approach that it preserves a central aspect of the classical picture of a particle by retaining the notion of a particle trajectory. This is the central point of difference between their approach and the framework of effects. Within the particle-effects framework the *important* incompleteness of quantum mechanics does not lie in its failure to recover particle effects within a propositional framework of dynamical properties, but in its failure to explain why particles have the peculiar character they do—why their representation is functional rather than propositional, with the consequence that particle-effects are knowable only with probability. A fruitful next step in the development of quantum foundations would be to address this question: the goal should not be to eliminate the incompleteness of quantum mechanics by the provision of an ontological story—not even one presented in terms of particle trajectories—but to explain it.

---

[9] This is not the most general formulation. Nevertheless, the consideration of contexts of the form *AB*, *AC*, … suffices to illustrate the possibility that an operator *A* may commute with *B* and *C*, but *B* and *C* may not commute with one another. It also suffices for the case in which contextuality coincides with non-locality and the observables qualify different component subsystems of a composite system of two particles.

*6. Conclusion*

So far as the problem of hidden variables is concerned, the framework of effects preserves the classical concept of reality in the sense of Pauli's characterization of it as one that assumes the possibility of a "detached observer." What it rejects is the adequacy of the propositional framework for understanding the peculiar probabilities of quantum mechanics: probabilities are assigned to effects—not to propositions belonging to elementary particles. The notion of an experimental set-up is fundamental to this account of the theory because particle-effects result from the interaction of particles with them. But *measurement* is not thereby made fundamental, since productions of particle-effects are not measurements. For the same reason human agency is not fundamental to the understanding of particle-effects, since experimental set-ups merely emulate naturally occurring conditions, albeit under the controlled circumstances characteristic of all experiments.

One consequence of the transition from propositions to effects is that it means taking the theory at face value in connection with its rejection of the construction of models within which one might picture elementary particles in terms of dynamical properties, after the manner familiar to us from classical physics. Interpretations of quantum mechanics have tended to be elaborations of different ways of filling out a story about the dynamical properties that is compatible with the theory; as such they have failed to address the fact that the probability theory implicit in quantum mechanics is itself indifferent to all such stories. This indeed is its novelty. For this reason, such interpretations must fail to capture what is really conceptually distinctive about theory, and what had always been emphasized by its great founders: the theory's radical departure from the usual spatio-temporal framework of classical physics and relativity. Our knowledge of elementary particles is restricted to their non-dynamical properties and their effects; and regarding the latter, our knowledge of them is inherently probabilistic. This is an evident rupture with our classical spatio-temporal picture of reality, but it involves no departure from our classical objectivist concept of reality. The present proposal for understanding the significance of the no hidden variable theorems may not position quantum mechanics as the great triumph of experimental science over realist metaphysics that some have imagined it to be, but it does present the theory as one of deep philosophical significance nonetheless: Quantum mechanics refutes the conception of the representational framework of elementary particles as a propositional one—as a totality of facts—whether this is elaborated within a realist or an anti-realist metaphysics.